\documentstyle[amsmath,multicol,prb,aps,graphicx]{revtex}

\def\prb{Phys. Rev. B }
\def\prl{Phys. Rev. Lett. }

\def\be{\begin{equation}}
\def\ee{\end{equation}}
\def\ba{\begin{eqnarray}}
\def\ea{\end{eqnarray}}

\def\LSCO{La$_{2-x}$Sr$_x$CuO$_4$ }

\def\YBCO{YBa$_2$Cu$_3$O$_{7-\delta}$ }
\def\124{YBa$_2$Cu$_4$O$_8$ }

\def\C60{A$_x$C$_{60}$ }

\begin{document}

\title{Weak-coupling phase diagrams of bond-aligned and diagonal 
doped Hubbard ladders}

\author{Shirit~Baruch and Dror~Orgad}

\address
{Racah Institute of Physics, The Hebrew University, Jerusalem 91904, Israel}

\date{\today}

\maketitle

\begin{abstract}

We study, using a perturbative renormalization group technique, the phase
diagrams of bond-aligned and diagonal Hubbard ladders defined as sections of 
a square lattice with nearest-neighbor and next-nearest-neighbor hopping.
We find that for not too large hole doping and small next-nearest-neighbor
hopping the bond-aligned systems exhibit a fully spin-gapped phase while
the diagonal systems remain gapless.
Increasing the next-nearest-neighbor hopping typically leads to a decrease
of the gap in the bond-aligned ladders, and to a transition into a gapped
phase in the diagonal ladders. Embedding the ladders in an antiferromagnetic
environment can lead to a reduction in the extent of the gapped phases.
These findings may suggest a relation between the orientation of hole-rich 
stripes and superconductivity as observed in \LSCO.

\end{abstract}


\begin{multicols}{2}

\section{Introduction}
\label{intro}

There is a growing number of experiments which suggest that the existence
of inhomogeneous electronic structures is a generic feature of the
hole-doped high-temperature superconductors. Often these structures take
the shape of quasi-one-dimensional hole-rich stripes running along various
directions in the copper-oxygen planes \cite{ourreview}. The relevance of
inhomogeneities in general and stripes in particular to the phenomenon of
high-temperature superconductivity is presently far from being a settled
issue.

Neutron-scattering experiments have produced clear evidence for a significant
correlation between signatures of incommensurate spin excitations in \LSCO
and the superconducting properties of this compound. Recently, it has been
demonstrated \cite{wakimoto04} that such low-energy incommensurate spin
excitations exist in the superconducting phase of \LSCO all the way up
to its overdoped boundary at $x=0.28$, where they disappear simultaneously
with superconductivity. On the underdoped side of the phase diagram static
magnetic stripe signatures exist both in the insulating spin-glass phase
\cite{diagstripes1,diagstripes2,diagstripes3} with $0.02\leq x\leq 0.05$
and in the underdoped superconductor
\cite{stripesLSCOsc1,stripesLSCOsc2,stripesLSCOsc3}
with $0.05\leq x\leq 0.12$. However, the transition from the insulator to
the superconductor at $x=0.05$ is accompanied by a rotation of the stripe
signal from the diagonal direction in the former to a bond-aligned signal
in the latter. Moreover, angle resolved photoemission spectroscopy (ARPES)
has revealed \cite{gapclose} an apparent coincidence between the collapse
of the gap along the nodal (diagonal) directions in $\vec{k}$-space and the
appearance of diagonal stripes in samples near the N${\rm \acute{e}}$el
boundary of the antiferromagnet at $x=0.02$.

Motivated by these observations we are interested in finding a possible 
correspondence between the geometry of stripes and their low-energy 
properties. However, studying the physics of stripes as self-organized 
systems in the strongly interacting copper-oxygen planes is a difficult 
task for which reliable analytical tools are still missing. Instead, we 
propose to investigate rudimentary models of stripes in the form of 
Hubbard ladders defined as bond-aligned and diagonal sections of a square 
lattice with nearest-neighbor $t$ and next-nearest-neighbor $t'$ hopping. 
In doing so we avoid the question of the mechanism which leads to the 
segregation of holes into stripes (on which there are different views  
\cite{psEK,castro1,whitestripes2d,geballe}), and 
assume that it takes place at energies higher than the energy scales which 
govern the physics of holes on the ladders. On the other hand, we will 
concentrate on a single ladder thus restricting the discussion to energies 
larger than any inter-stripe coupling (such as inter-stripe tunneling), 
which is assumed small. While the degree to which holes are restricted  
to move along stripes in the real systems is not well understood it is 
clear that the limit of total confinement, assumed in the ladder models,
is an idealization. To probe the sensitivity of the results to this point 
we consider, for each of the two geometries mentioned above, both 2 and 
3-leg ladders.  

Arguably, the most severe shortcoming of the present study, with respect to 
its applicability to the physics of stripes in the cuprates, is the fact that 
the interaction between electrons on the ladders is assumed weak and treated 
using a perturbative renormalization group (RG) technique. Nevertheless, 
previous numerical studies\cite{noackprl,noackphys,kimura,3leg,compare,zigzag}
have shown that many of the qualitative features uncovered
by perturbative analysis of Hubbard ladders survive the limit of
strong coupling where the emergent energy scales are of the order of
the exchange interaction $J$. This fact suggests that the gross features 
which we discover in the phase diagrams of the models considered here may 
likewise extend into the strong coupling regime. Of course, this needs to 
be verified by explicit calculations. 

Our principal results are presented in Figs. \ref{bond2LLFig} -
\ref{diag3LLFig}. They contain the phase diagrams of the ladders
as a function of the relative strength of next-nearest-neighbor
hopping $t'/t$ and the average electronic occupation per site $n$.
The various phases are labeled according to the number of charge
and spin modes \cite{bf2ll,bfnll,classification} that remain
gapless in the presence of interactions. We find a distinct
difference between the behavior of bond-aligned and diagonal
ladders for small values of $t'/t$ and doping levels in the range
of $0.5\leq n \leq 1.5$. In this region the bond-aligned ladders
are maximally-gapped with only one remaining gapless charge mode
associated with global gauge and translational invariance
\cite{classification} (at commensurate fillings this mode may be
gapped as well). This finding extends previous results for such
models obtained both from weak-coupling treatments
\cite{bf2ll,bfnll,classification,schulz,spingap,dhlee,optimal} and
numerical calculations at strong-coupling
\cite{noackprl,noackphys,kimura,3leg,compare}. Conversely, the
diagonal ladders remain gapless under similar conditions. Such
behavior has been known for the 2-leg ``zig-zag'' ladder
\cite{fabrizio,zigzag,zigzag01} and our results demonstrate that it
persists also in the case of the diagonal 3-leg ``diamond''
ladder.

Increasing the value of $t'/t$ in bond-aligned systems, which exhibit a
gapped phase for zero $t'$, leads to a decrease of the gap. In the diagonal
ladders similar variation of the next-nearest-neighbor hopping results in a
transition from a gapless to a gapped phase above a certain critical value.
Further increase of $t'/t$ enhances the gap, which reaches a maximum and
then decreases. 

In order to gain some insight into the effects of a magnetically ordered 
background on the physics of ladders we replace the non-interacting spectrum 
of the ladder by the bound-states spectrum of a linear defect in an 
antiferromagnetic two-dimensional lattice. In this way we include on a crude 
mean-field level the interactions between the stripe electrons and the 
surrounding spins. The interactions between electrons on the defect are 
still treated perturbatively. We find that the changes in the spectrum typically
have little effect on the overall structure of the phase
diagrams, as long as the electronic states are reasonably localized on
the defects. When changes do occur they often tend to reduce the extent of
the gapped phases.

Notwithstanding the above mentioned limitations of our study, we note that 
our findings, and the experimental congruity between stripes orientation
and superconductivity, conform with the notion that the origin of the
superconducting gap is to be found in the gap which develops on the
quasi-one-dimensional stripes \cite{spingap}. The results may also provide
an explanation, within the stripes picture, for the dichotomy between
gapped and gapless behavior in $\vec{k}$-space, as revealed by ARPES. We
elaborate on these points
and raise few difficulties related to such
interpretations in the discussion section.

\section{Stripes as Ladders}
\label{ladders}

\subsection{Models and methods}
\label{laddermodel}

\begin{figure}[h]
\includegraphics[angle=0,width=3.2in]{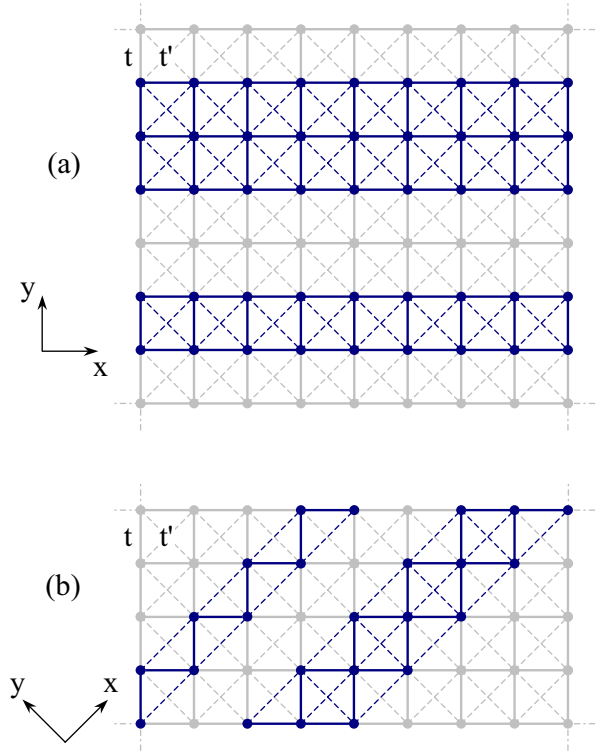}
\caption{2 and 3-leg ladders on the background lattice.
(a) bond-aligned ladders (b) diagonal ladders.}
\label{laddersFig}
\end{figure}

Consider the Hubbard model on a square lattice with
nearest-neighbor $t$ and next-nearest-neighbor $t'$ hopping. We
begin by modeling a single stripe as a quasi-one-dimensional
section consisting of $N$ connected infinite chains, {\it i.e.} an
N-leg ladder, on this lattice. In this paper we will concentrate
on the minimal models for bond-centered and site-centered stripes
as 2-leg and 3-leg ladders, respectively. We will consider two
geometries for the ladders: bond-aligned and diagonal. In the
first the ladder lies along the nearest-neighbor bonds while in
the second it extends in the direction of the
next-nearest-neighbor bonds (See Fig. \ref{laddersFig}). In both
cases we assume open boundary conditions in the transverse
direction.

The bond-aligned N-leg ladder is described by the
Hamiltonian $ H = H_0^{\textit{bond}} + H_U$
\ba
H_0^{\textit{bond}}&=& \sum_{i=1}^N\sum_{x}
\sum_{\sigma=\uparrow,\downarrow} \bigg\{ \nonumber \\
&-&t[c^{\dag}_{i,x,\sigma} c^{\vphantom\dag}_{i+1,x,\sigma}
+ c^{\dag}_{i,x,\sigma} c^{\vphantom\dag}_{i,x+1,\sigma} + h.c. ]
\nonumber\\
&-&t' [ c^{\dag}_{i,x,\sigma} c^{\vphantom\dag}_{i+1,x-1,\sigma}+
c^{\dag}_{i,x,\sigma} c^{\vphantom\dag}_{i+1,x+1,\sigma}+ h.c.]\bigg\} \; ,
\nonumber \\
\\
\label{Hubbardterm}
H_U&=& U \sum_{i=1}^N\sum_{x}: n_{i,x,\uparrow}n_{i,x,\downarrow}: \; ,
\ea
where $ c^{\vphantom\dag}_{i,x,\sigma}$ annihilates a spin
$\sigma$ electron at site $x$ on chain $i$ and
$n_{i,x,\sigma}=c^\dagger_{i,x,\sigma}c^{\phantom{\dagger}}_{i,x,\sigma}$.
For the bond-aligned ladders we take as unit length the distance between
nearest neighbors.
The on-site Hubbard interaction is assumed weak $U\ll t$.

The diagonal N-leg ladder is described by the Hamiltonian
$H=H_0^{\textit{diag}} + H_U$ where
\ba
H_0^{\textit{diag}}&=&
\sum_{i=1}^N\sum_{x}\sum_{\sigma=\uparrow,\downarrow}\bigg\{ \nonumber \\
&-&t[c^{\dag}_{i,x,\sigma}c^{\vphantom\dag}_{i+1,x+\frac{1}{2},\sigma} +
c^{\dag}_{i,x,\sigma}c^{\vphantom\dag}_{i+1,x-\frac{1}{2},\sigma} + h.c.]
\nonumber\\
&-&t' [ c^{\dag}_{i,x, \sigma} c^{\vphantom\dag}_{i,x+1,\sigma} +
c^{\dag}_{i,x, \sigma} c^{\vphantom\dag}_{i+2,x,\sigma} + h.c.] \bigg\} \; .
\ea
For this model we choose the next-nearest-neighbor distance as our unit
of length along the $x$ direction. This means that $x$ runs over the
integers on odd chains and over the integers shifted by 1/2 on even chains.

We study the above models in the weak-coupling limit using RG techniques.
Much of the analysis is identical to the one used in previous studies of
Hubbard ladders \cite{bf2ll,bfnll,classification,schulz,spingap,dhlee,optimal}
and consequently we outline here only the basic steps of the procedure,
relegating some of the remaining details to the appendix.

Since the interaction term is treated as a perturbation the first step is the
diagonalization of the quadratic part of the Hamiltonian via the
transformation
\be
\label{transformation}
c_{i,x,\sigma}=\sum_{b=1}^N\sum_{k_x=-\pi}^\pi
e^{ik_xx}S_{ib}\psi_{b,\sigma}(k_x) \; ,
\ee
where the matrix $S$ depends on the specific model. The resulting free
band structure is then filled up to the chemical potential (we assume
zero temperature throughout the paper) which is determined by the averaged
electronic occupation per site $n$. Each band that crosses the chemical
potential gives rise to $N_b$ symmetric pairs of Fermi points, where more
than one pair may exist for a single band, see Fig. \ref{separation} for an
example. A linearized chiral electronic branch is associated with every
Fermi point. In order to have positive definite Fermi velocities we assign
left (right) moving fields to Fermi points at which the spectrum is a
descending (ascending) function of $k_x$. When the expansion of the
band field operators in terms of these chiral fields
\be
\label{opexpansion}
\psi_{b,\sigma}(x)\sim \sum_{j_b=1}^{N_b} e^{i \eta_{j_b}|k_{F_{j_b}}|x}
\psi_{j_b,\sigma}^R(x)+ e^{-i \eta_{j_b}|k_{F_{j_b}}|x}
\psi_{j_b,\sigma}^L(x) \; ,
\ee
with $\eta={\rm sign}(k_F)$ at the right moving point, is plugged into the
Hubbard term, Eq. (\ref{Hubbardterm}), a host of interaction terms emerges.
Typically only forward- and Cooper-scattering are allowed by momentum
and spin conservation \cite{bf2ll,bfnll}, but for appropriate values of
the parameters other momentum-conserving ``special'' processes and 
Umklapp terms appear as well (See Figs. \ref{bond2LLFig},\ref{diag2LLFig} 
and the appendix).

In the first stage of the renormalization process the RG equations for
the various coupling constants (as given in the appendix) are integrated
numerically until $\l$, the logarithm of the cut-off in units of its
initial value, approaches a point $\l^*$ where one or more of the
couplings grows as $(l-l^*)^{-\gamma}$ with $\gamma\leq 1$. At this
scale the flow has reached the vicinity of a strong-coupling fixed point.
The most divergent couplings in this region tend to freeze certain modes
and open a gap in their fluctuation spectrum.
The pinned modes may be identified via bosonization and the gap scale
associated with them may be related to the original bandwidth $E_F$
according to
\be
\label{gapscale1}
\Delta_1\approx E_F e^{-l^*} \; .
\ee

To obtain the behavior of the system on scales larger than $l^*$ one
needs to carry out a second RG stage in which the relevance of the weakly
or non-diverging couplings of the first stage is assessed near the new
strong-coupling fixed point \cite{classification,optimal}. This is done
by calculating their scaling-dimension $\delta$ while freezing the gapped
modes. In case some of these couplings become relevant the flow is directed
towards a new fixed point in which additional modes are gapped with a
typical scale of
\be
\label{gapscale2}
\Delta_2\approx \Delta_1 g_*^{1/(2-\delta)} \; ,
\ee
where $g_*$ is the dimensionless strength of the residual interactions at
the end of the first RG stage \cite{classification,optimal}.
In principle one should repeat the process until a stable fixed point is
reached. In our case we find that if new relevant couplings do occur in the
second stage, they take the system into the maximally allowed gapped phase.

\subsection{Results}
\label{laddersresults}
\begin{figure}[h]
\includegraphics[angle=0,width=3.2in]{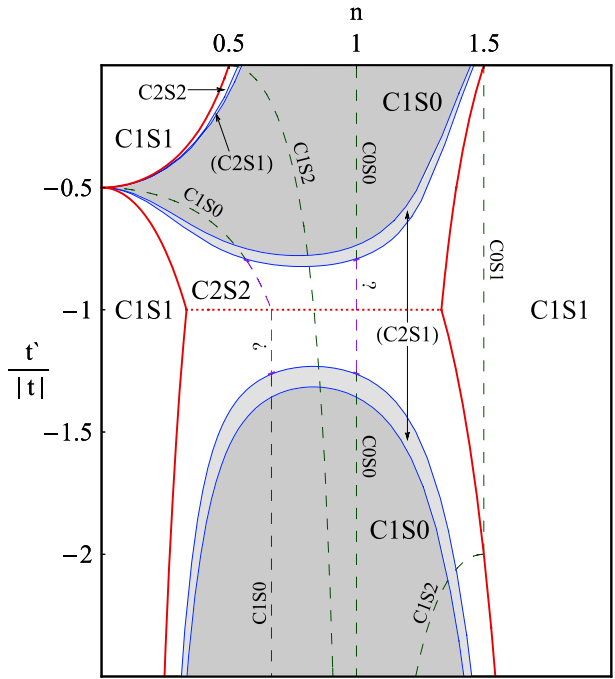}
\caption{The phase diagram of the 2-leg bond-aligned ladder. The system
remains gapless in the light regions and exhibits the maximally gapped
C1S0 phase in the shaded parts of the diagram. In the darker shaded regions
this phase is obtained at the end of the first RG stage while it is reached
only at the end of the second stage inside the lightly shaded domains. In
the latter case we indicate in parentheses the nature of the phase at the
end of the first stage. Umklapp and ``special'' processes are allowed on
the dashed lines.
The number of non-interacting Fermi points changes by 2 across the thick
solid lines and the Fermi velocity associated with the pair vanishes there.
On the dotted line a flat band occurs. Consequently, our results are
unreliable in the vicinity of these lines.}
\label{bond2LLFig}
\end{figure}

\begin{figure}[h]
\includegraphics[angle=0,width=3.2in]{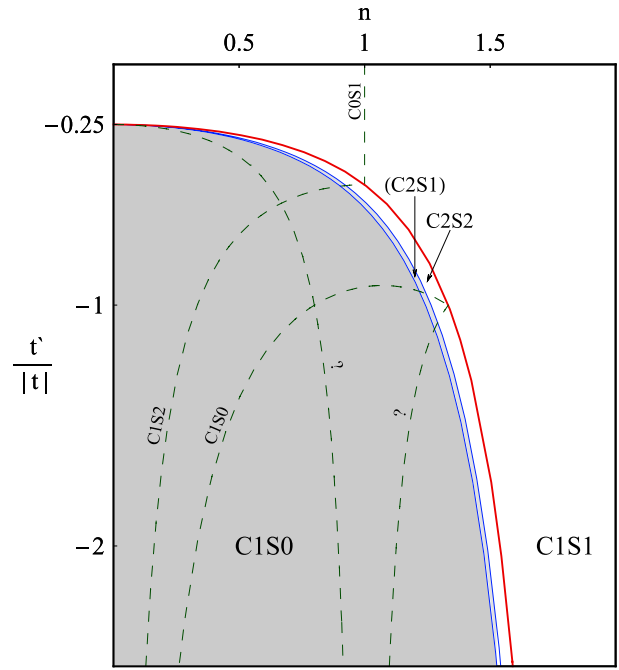}
\caption{The phase diagram of the 2-leg diagonal ladder.}
\label{diag2LLFig}
\end{figure}

\begin{figure}[h]
\includegraphics[angle=0,width=3.2in]{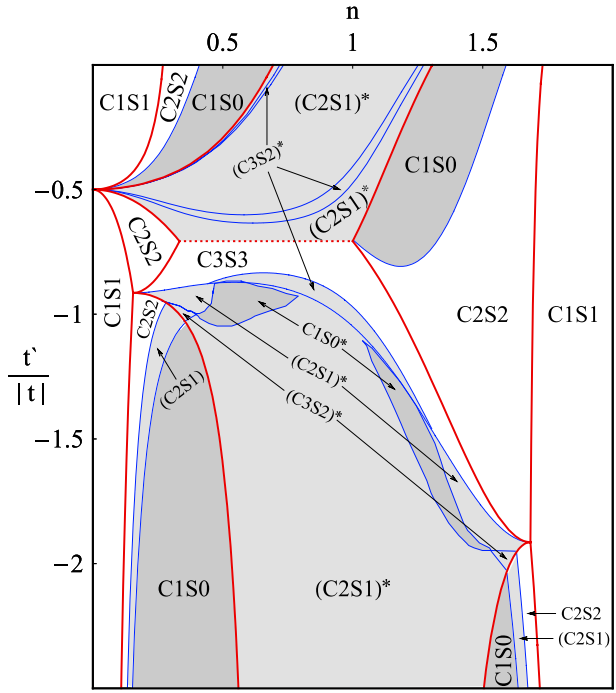}
\caption{The phase diagram of the 3-leg bond-aligned ladder. To distinguish
between gapped phases which originate from a C2S2 phase, and those coming
from a C3S3 phase, we mark the latter by an asterisk.}
\label{bond3LLFig}
\end{figure}

\begin{figure}[h]
\includegraphics[angle=0,width=3.27in]{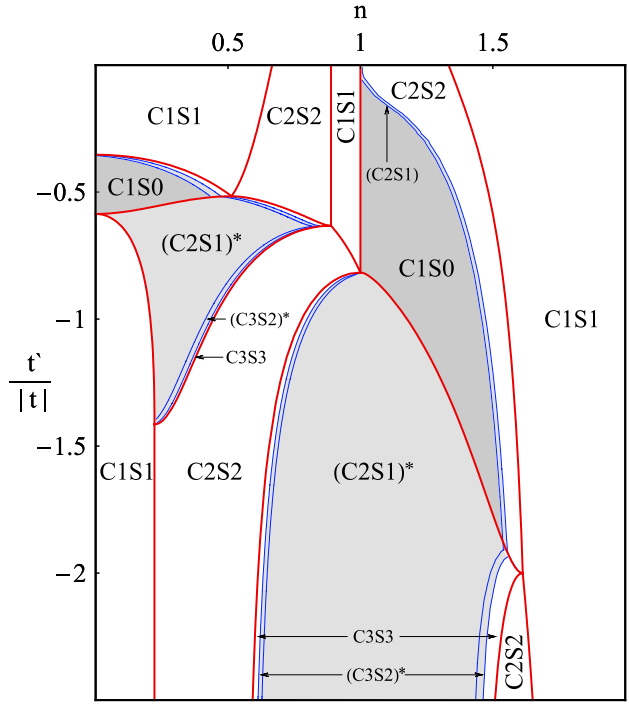}
\caption{The phase diagram of the 3-leg diagonal ladder.}
\label{diag3LLFig}
\end{figure}

\vspace{-0.4cm}
Our results for the ladder models are summarized in the phase
diagrams as presented in Figs. \ref{bond2LLFig} - \ref{diag3LLFig}.
To facilitate comparison with the cuprates we concentrate on the case
$t'<0$\cite{mcmahan,andersen}. Since the model is defined on a bipartite
lattice the spectrum is invariant under $t\rightarrow -t$. Moreover, the
particle-hole transformation
$c_{x,y,\sigma}\rightarrow c^\dagger_{x,y,\sigma}$ maps the original
Hamiltonian on itself with $t\rightarrow -t$, $t'\rightarrow -t'$ and
changes the average electronic density according to $n\rightarrow 2-n$.
One can therefore obtain the $t'>0$ half of the diagrams by reflecting
them with respect to the $t'=0$ and the $n=1$ lines.

We use the C$m$S$n$ classification scheme\cite{bf2ll,bfnll,classification}
wherein a phase is characterized by the number ($m,n$) of its gapless charge
and spin modes. In the diagrams we shaded the regions in which the systems
end up in the maximally gapped phase (C1S0) allowed by global gauge and
translational invariance\cite{classification}. In the lightly shaded areas
this phase is obtained only at the end of the second RG stage and the
nature of the phase at the end of the first stage is shown in parentheses.
More information on the character of the massive fields, and the relevant
operators which give them their masses, is detailed in the appendix.
We find that whenever the C1S0 phase is not reached the systems retain all
of their non-interacting gapless modes, as indicated in the clear parts of
the diagrams.

For the 2-leg ladders we have also considered the possibility of Umklapp
scattering and momentum-conserving ``special'' processes which exist for 
particular values of the Fermi wave-vectors. The dashed lines in
the diagrams correspond to parameters for which such processes are allowed.
In all cases we find that the nature of the resulting phases, as indicated
next to the lines, is determined by the end of the first RG stage.
Along the sections marked by ``?'' the relevant terms of the first
RG stage tend to pin pairs of canonically conjugated fields.
Consequently, we were unable to determine the fate of the system in
this eventuality.

Our renormalization procedure relies on the linearization of the
non-interacting spectrum around the Fermi points. It therefore fails
whenever a pair of Fermi points resides at an extremum of the
band structure. This happens along the thick solid lines in the diagrams.
Across each of these lines the number of Fermi point pairs changes by 1
(or 2 in the case of a ``W'' shaped band). For the bond-aligned ladders
a flat band arises along the dotted lines in their diagrams while for the
3-leg diagonal ladder such a band occurs for $t'=0$ and $2/3<n<4/3$.
Our results are unreliable in the vicinity of these lines.

Our most important finding is a correlation between the ladder orientation
and the nature of its phase at small values of $t'/t$. The bond-aligned
ladders exhibit the gapped C1S0 phase over a wide range of electronic filling
$0.5\lesssim n\lesssim 1.5$. The diagonal ladders on the other hand remain
gapless, except for a small region near $n=1$ in the case of the diagonal
3-legged system.

Bond-aligned ladders with $t'=0$ have been analyzed in the past using
perturbative RG\cite{bf2ll,bfnll,classification,schulz,spingap,dhlee,optimal}
and in most cases were found to be in the C1S0 phase\cite{comment}. This
conclusion was further supported by density matrix renormalization group
(DMRG) calculations at strong-coupling
\cite{noackprl,noackphys,kimura,3leg,compare}. Our results demonstrate
the robustness of this phase in the presence of not too large
next-nearest-neighbor hopping. Our results also agree with the gapless
behavior found in a previous DMRG study of the diagonal 2-leg
ladder\cite{zigzag} and add information about this system at commensurate
fillings. We are unaware of past treatments of the 3-leg diagonal ladder,
especially in the strong interaction regime. It is therefore unclear whether
the gapless phase found by us in this model survives beyond the limit
of weak-coupling. It is, however, interesting to note that at half filling
and for $t'=0$ and $U\rightarrow\infty$, the diagonal 3-leg ladder is
equivalent to an alternating spin-1 and spin-1/2 chain, which is known to
be gapless\cite{alternating}.

\vspace{-0.37cm}

One may estimate the gap scale in the gapped phases from Eqs.
(\ref{gapscale1},\ref{gapscale2}). Typically we find that the gap changes
smoothly across boundaries between adjacent extended C1S0 phases, regardless
of the RG stage at which they become fully gapped. One exception is the
3-leg bond-aligned ladder at small $t'/t$ where the gap increases
considerably in the region around $n=1$ due to a much smaller $l^*$
(which, in passing, we note scales as $t/U$). The gap is very small in the
sliver phases.

\vspace{0.07cm}
\begin{figure}[h]
\includegraphics[angle=0,width=3.2in]{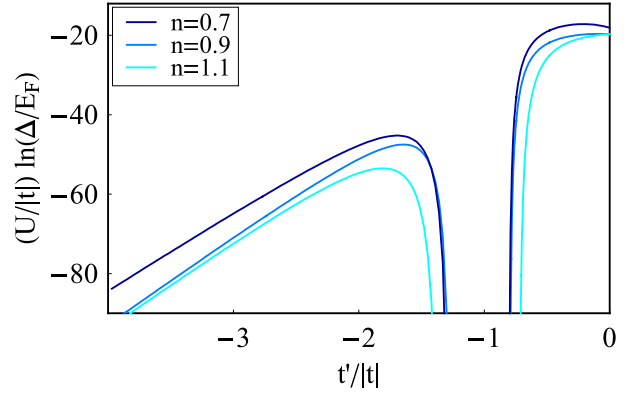}
\caption{Gap scale in the 2-leg bond-aligned ladder.}
\label{bond2LLgap}
\end{figure}

\vspace{0.07cm}
\begin{figure}[h]
\includegraphics[angle=0,width=3.2in]{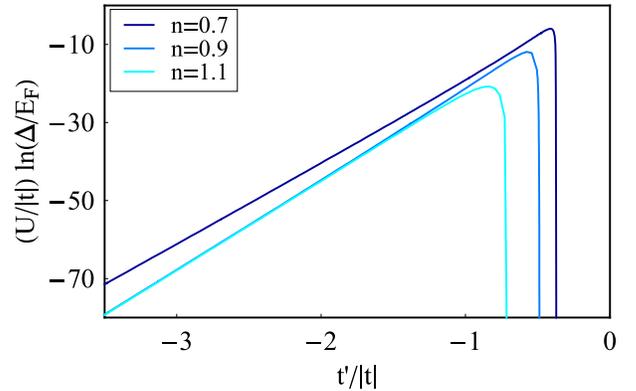}
\caption{Gap scale in the 2-leg diagonal ladder.}
\label{diag2LLgap}
\end{figure}

The dependence of $\Delta$ on $t'$ for the 2-leg ladders is presented
in Figs. \ref{bond2LLgap},\ref{diag2LLgap}. Increasing the magnitude of the
next-nearest-neighbor hopping results in a decrease of the gap in
the bond-aligned system. On the contrary, the diagonal ladder, which is
gapless for $t'=0$, develops a gap above a critical density-dependent
value of $t'$. Further increase of the hopping amplitude causes the gap to
grow, reach a maximum and then decrease. We observe a similar behavior
also in the 3-leg ladders.

\section{Stripes as defects in an antiferromagnetic background}
\label{defects}

\subsection{Model}
\label{defectmodel}

Since the cuprate high-temperature superconductors are born out of
parent antiferromagnets one expects to find slow fluctuations of a
collective field representing the local staggered magnetization.
Stripes may be thought of as defects in this field along which
holes tend to segregate. In the previous section we have assumed
that the confinement of the holes to the stripes is extremely strong
such that their coupling to the antiferromagnetic environment may be
neglected. Here we would like to relax this constraint. To do so we
consider a model of non-interacting electrons on the square lattice
interacting with a static staggered field
\ba
\label{Hdefect}
H_0^{\textit{defect}}&=&\sum_{x,y}\sum_{\sigma=\uparrow,\downarrow}
\bigg\{\nonumber\\
&-&t[c^{\dag}_{x,y,\sigma}c^{\vphantom\dag}_{x+1,y,\sigma} +
c^{\dag}_{x,y,\sigma}c^{\vphantom\dag}_{x,y+1,\sigma} + h.c. ] \nonumber\\
&-&t' [c^{\dag}_{x,y,\sigma} c^{\vphantom\dag}_{x-1,y+1,\sigma}
+c^{\dag}_{x,y,\sigma} c^{\vphantom\dag}_{x+1,y+1,\sigma} + h.c.]\nonumber\\
&+& \sigma(-1)^{x+y}m(x,y)c^{\dag}_{x,y,\sigma}
c^{\vphantom\dag}_{x,y,\sigma} \bigg\} \; ,
\ea
where $c^{\vphantom\dag}_{x,y,\sigma}$ annihilates a spin $\sigma$
electron at site $(x,y)$ on the lattice (as defined in the coordinate
system of Fig. \ref{laddersFig}a.) The field configuration
is taken to represent defects with the same geometries as the ladders
studied in Section \ref{ladders}
\be
m(x,y)=\begin{cases}
m &\text{left of ladder} \\
0 &\text{on the ladder} \\
\pm m &\text{right of ladder} \; . \\
\end{cases}
\ee
We consider both in-phase ($+m$) and anti-phase ($-m$)
domain walls.

\begin{figure}[h]
\includegraphics[angle=0,width=3.1in]{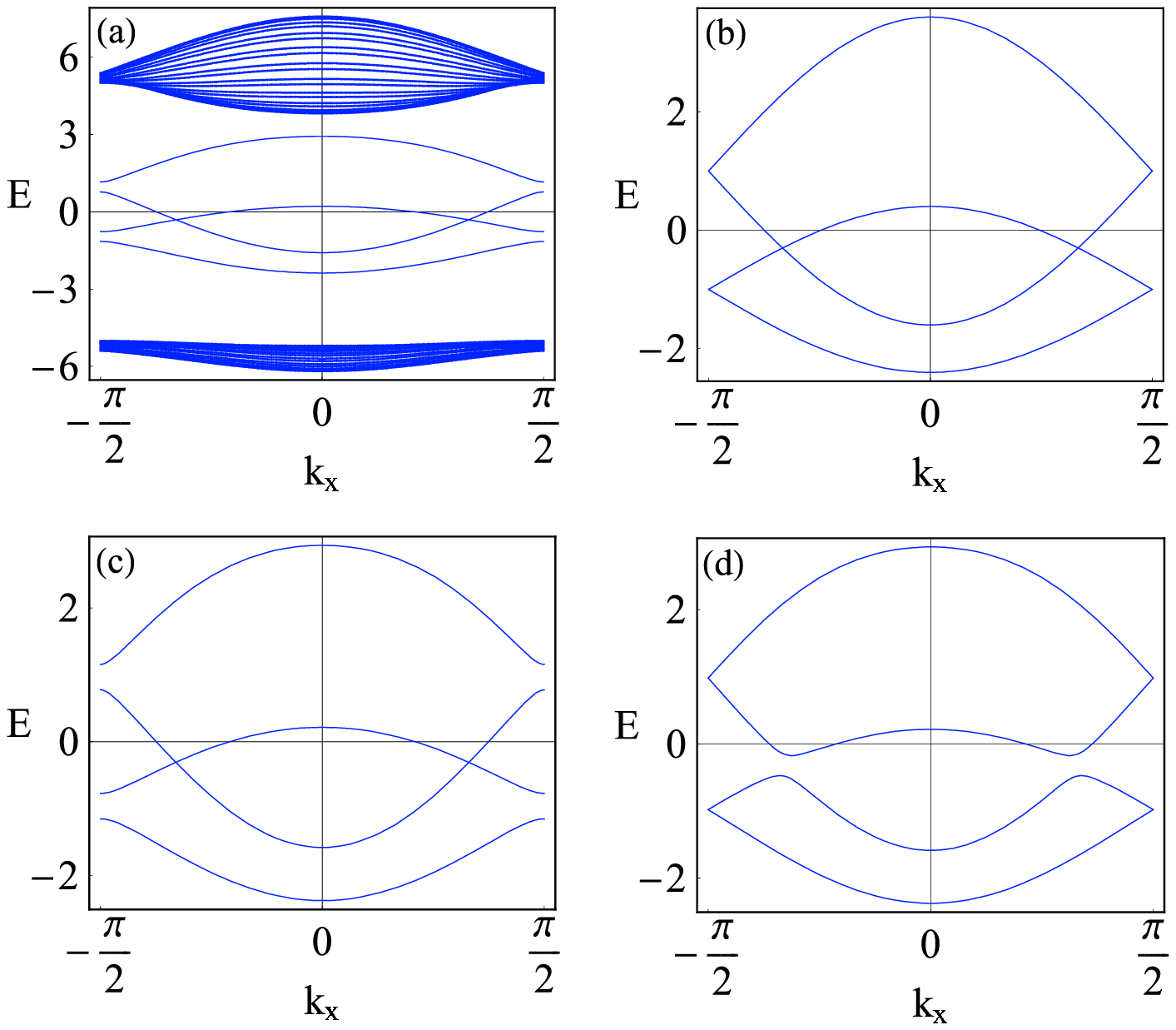}
\caption{The band structure of the 2-leg bond-aligned defect for $t'/t=-0.3$.
(a) The full spectrum including the bands of extended states for the case of
an anti-phase defect with $m/t=5$, (b) the localized bands for infinite $m$,
(c) the localized bands of the anti-phase defect with $m/t=5$, (d) the
localized bands of the in-phase defect with $m/t=5$.}
\label{separation}
\end{figure}

In the absence of a defect, {\it i.e.} $m(x,y)=m$, the spectrum exhibits 
two bands of extended Bloch states whose energies
$E=-4t'\cos(k_x)\cos(k_y)\pm\sqrt{4t^2[\cos(k_x)+\cos(k_y)]^2+m^2}$
are separated by an energy gap of order $m$ for large enough magnetization.
Introducing a domain wall into the periodic background leads to the
appearance of mid-gap states, see Fig. \ref{separation}. These states
are localized in the transverse direction to the wall\cite{stripes}
over a distance which typically scales as $m^{-1}$. We therefore
expect their energies to approach, in the limit $m\rightarrow\infty$,
the non-interacting energies of the corresponding ladder.

In the following we assume that $m$ is large enough, such that varying the
density of the system away from half filling is achieved by doping
electrons or holes solely into the mid-gap bands. Consequently, we consider
a reduced model whose non-interacting spectrum is composed of these
localized mid-gap bands only. This band structure expresses, albeit on
a coarse mean-field level, the effects of interactions between
the electrons on the defect and the electrons comprising the
antiferromagnetic background. In the cuprates these interactions are 
expected to be of similar magnitude to the interactions between the  
electrons moving along the stripes. However, in our perturbative 
treatment the latter are assumed much weaker. While unjustified when 
modeling real stripes this assumption enables us to treat the 
interactions on the defect in a controlled manner and gain some 
insight into the effects of the coupling between stripes and their  
antiferromagnetic environment.  

To make contact with the ladder models of the previous
section we relate the electronic operators on the defect sites to the
creation operators of the localized states. The result is analogous
to Eq. (\ref{transformation}) with two differences. First, the exact
expansion of the site operators generally includes components along
extended states. Such components are projected out and the remaining
sum is properly normalized. As long as the mid-gap states are well
localized on the defect the projected expansion remains a faithful
representation of the site operators on the stripe. Secondly, for an
$N$-leg bond-aligned defect the staggered field induces a doubling of
the unit-cell and of the number of localized bands. Accordingly, in
Eq. (\ref{transformation}), $x$ becomes the unit-cell coordinate and
$i$ stands for the leg number together with the site index within the
unit-cell. The sum over the band index then runs from 1 to $2N$
and the sum over the momentum along the defect extends from $-\pi/2$
to $\pi/2$. This effect is absent in the case of diagonal stripes.
On the other hand, and in contrast to the bond-aligned systems, the spin
degeneracy of the bands is lifted whenever the spin texture is symmetric
on opposite sides of a diagonal defect [this happens in anti-(in-) phase
defects with even(odd) $N$].

From this point onward the analysis of the interactions
between electrons on the defect, as given by the Hubbard term
Eq. (\ref{Hubbardterm}), follows the same course described for the
ladder models. However, owing to the absence of spin SU(2) symmetry
in this case, we are compelled to consider, separately, scattering
processes in different spin channels.\cite{dhlee} The resulting RG 
equations will be given elsewhere.

\begin{figure}[h]
\includegraphics[angle=0,width=3.2in]{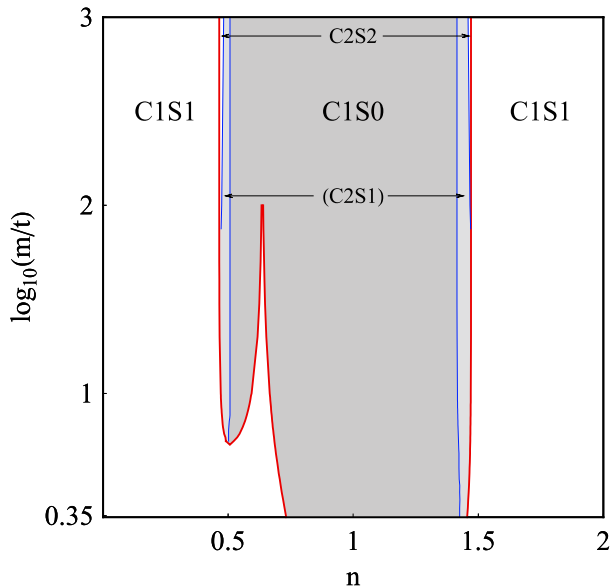}
\caption{The dependence of the phase diagram of an anti-phase 2-leg
bond-aligned defect on the magnetization $m$. The results are for the
case $t'/t=-0.1$.}
\label{antibondt01Fig}
\end{figure}

\subsection{Results}
\label{defectsresults}

In Fig. \ref{antibondt01Fig} we present the phase diagram of the
anti-phase bond-aligned 2-leg defect as a function of $m$ and $n$
for a fixed small value of $t'$. We already referred to the fact that
identifying the mid-gap sector of the spectrum with the electronic
states of a stripe makes sense as long as these states are well
localized on the defect. Since the localization length becomes
larger with decreasing magnetization or increasing next-nearest-neighbor
hopping it imposes a limit on the minimal $m$ and maximal $t'/t$
which may be studied using this model. This constraint
sets the lower bound of the magnetization axis in the figure.

As can be seen lowering the modulation depth of the antiferromagnetic
environment leads eventually to a reduction in the extent of the gapped
phase. The corresponding diagram for the analogous in-phase system, on
the other hand, shows no significant changes in the phase boundaries down
to the lowest value of $m$. A similar result is obtained for the 3-leg
bond-aligned defects, with the exception that it is the in-phase system
whose gapped phase is depleted at small $m$ while the diagram of the
anti-phase defect remains largely unaffected. We are therefore led to
the conclusion that a larger sensitivity to changes in $m$ exists when
spins across the defect are aligned.

\begin{figure}[h]
\includegraphics[angle=0,width=3.2in]{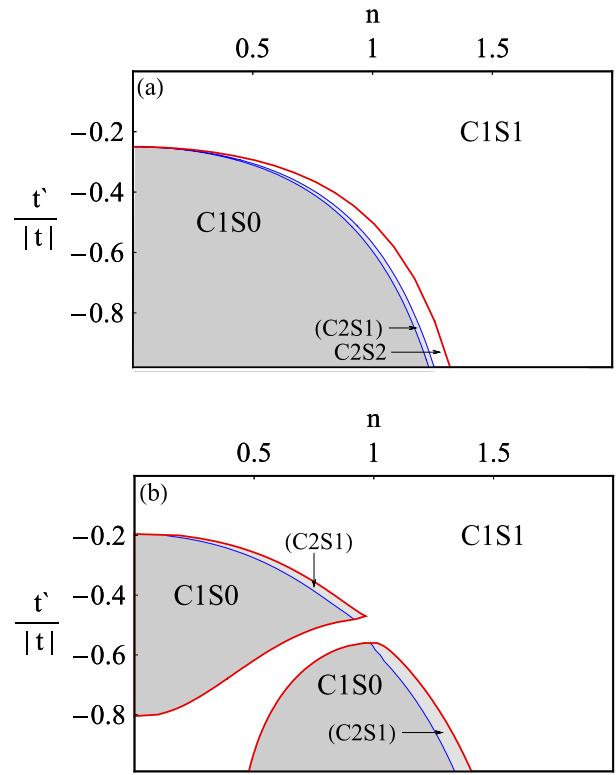}
\caption{A comparison between the phase diagrams of (a) the 2-leg diagonal
ladder and (b) the in-phase 2-leg diagonal defect with $m=3.5$.}
\label{indiag2LLm35Fig}
\end{figure}

The changes in the phase diagrams are initiated by deformations in
the non-interacting band-structure as $m$ is reduced. For infinite
$m$ the mid-gap spectrum of the defect is identical to that of the
corresponding ladder. For finite magnetization some of the degenerate
points in the spectrum of the ladder separate by mini-gaps. When the
background spin texture is symmetrical with respect to the defect
the degeneracy at $k_x=\pi/2$ is lifted. When it is asymmetrical
the bands separate at $k_x$ values away from the edge of the Brillouin
zone, see Fig. \ref{separation}. When a separation occurs,
the result may be the removal of two Fermi points from the system
(an example appears in Fig. \ref{separation}c around $E=-1$).
Such an event typically leads to a transition from a gapped to
a gapless phase. In the symmetrical case, this scenario is more
frequent, a fact which explains why it is more sensitive to variations
in $m$. Opposite occurrences where a band deforms from a ``U''
to a ``W'' shape, with the result of the appearance of an additional
pair of Fermi points, are also possible but are less common.
These changes lead to fragmentation of the phase diagrams of the defects.
Usually gapped phases are being penetrated by extensions of gapless
phases, but we also encounter pockets or slivers of strong
(first RG stage) C1S0 phase inside regions which are gapless or weakly
gapped in the ladder system.

We already mentioned that the spin degeneracy of the spectrum is lifted
in the anti-phase 2-leg and the in-phase 3-leg diagonal defects. In
these cases the RG analysis indicates that the coupling constants
do not flow and the systems remain in their gapless non-interacting
phases. The spectra of the related systems with anti-symmetric
spin configurations across the defects are still spin-degenerate and
contain similar openings of mini-gaps as the bond-aligned defects.
Consequently, their phase diagrams exhibit a more fragmented behavior,
as demonstrated by Fig. \ref{indiag2LLm35Fig}.

We find that lowering the magnitude of the staggered field diminishes
the extent of the C2S2 phase throughout the phase diagrams. This effect
is especially significant in the case of the 3-leg anti-phase diagonal
defect where it leads to a reduction in the extent of the gapless phase in
small t'/t, narrowing it down to a region centered around half filling.

\section{Discussion}

The results presented in this paper were derived for weak interaction
strength $U$ where all gap scales are exponentially small. It is unclear 
to what extent they represent the properties of ladders at strong coupling. 
However, based on past experience coming from numerical studies 
\cite{noackprl,noackphys,kimura,3leg,compare,zigzag}, it seems that many 
of the qualitative features of doped Hubbard ladders, such as the nature of  
their low-energy phases, are robust against changes in the interaction 
strength. In the following we assume that such continuity holds also for 
our results and, keeping in mind the other limitations of modeling 
hole-rich stripes as doped ladders, comment on their possible relevance 
to the physics of the cuprate high-temperature superconductors.
 
By now a large body of experimental data points to the fact that
inhomogeneous electronic structures are prevalent in the cuprates
\cite{ourreview}. There is no consensus, however, on the role played
by such inhomogeneities in the formation of high-temperature
superconductivity and the other unique features of these compounds. 
One view, identifies the scattering of quasiparticles from the 
critical charge fluctuations associated with a stripe quantum critical 
point, as responsible for the anomalous electronic properties, such as 
the pseudogap in the underdoped regime \cite{singular,qcpscenario}. 
Another view, stipulates that the spin-gap, which develops on stripes, is 
the source of the pseudogap \cite{spingap}. The opening of this gap leads 
to intra-stripe power-law superconducting correlations, 
which turn eventually, via inter-stripe Josephson pair-tunneling, 
into global long-range superconducting order.

For stripes to play an essential role in the physics of the cuprates 
they must exist, either as static order or as fluctuating correlations, 
above the pseudogap crossover and the superconducting transition 
temperatures. Moreover, if the second point of view, just mentioned, 
is correct, these stripes must support a gap in order for superconductivity 
to be achieved.

As indicated in the introduction \LSCO is a system where the existence
of stripes is very well established over a wide range of hole-dopings
and temperatures. As such it fulfills the first requirement. Our finding
that under generic conditions weakly interacting diagonal ladders are 
gapless while bond-aligned ladders possess a spin-gap, offers some support 
to the hypothesis that a similar difference between the low-energy 
properties of bond-aligned and diagonal stripes is behind 
the coincidence of the insulator-superconductor transition and the rotation
of the stripes direction in this compound\cite{machida}. It also supports the identification
of bond-aligned stripes as the origin of the gapped ARPES spectrum
around $(\pi,0)$ (and symmetry related points) in $\vec{k}$-space, and the
association of the gapless Fermi surface at $(\pi/2,\pi/2)$ with diagonal
stripes.

One may raise, however, difficulties associated with this picture.
The first concerns signs of gapped components in the spin-glass phase
of \LSCO with $x\leq 0.05$. ARPES measurements have found a distinctive
``knee'' in the energy distribution curves at $(\pi,0)$ in $x=0.05$
samples\cite{arpessit,arpeslsco,arpesarc}. This feature appears at a
binding energy of about 0.1 eV and becomes more prominent and less gapped
with increased doping levels. It is consistent, within experimental
uncertainties involving the doping, with the bond-aligned incommensurate
signal observed in elastic neutron-scattering experiments\cite{diagstripes3}.
This signal coexists with incommensurate diagonal peaks for
$0.056\leq x \leq 0.07$ and grows stronger with $x$. There is, however, some
evidence coming from ARPES\cite{arpessit,arpeslsco,arpesarc} that a
similar but faint gapped band exists also in $x=0.03$ samples for
which no corresponding neutron-scattering signal has been reported.

A more serious objection, perhaps, may be raised in connection to the
existence of a gapless diagonal component well inside the superconducting
regime. A peak in the ARPES spectrum near $(\pi/2,\pi/2)$ and at the Fermi
energy exists in all samples with $0.03\leq x \leq 0.15$
(and beyond)\cite{arpesarc,arpeslsco}. Its spectral weight intensifies
upon entering the superconducting region and continues to grow at least
until optimal doping\cite{arpesarc}. Although the above mentioned coexistence
of bond-aligned and diagonal incommensurate  neutron-scattering peaks
might explain this behavior near the insulator-superconductor transition,
it does not offer an explanation at higher doping levels. One possible
solution to the problem is that diagonal stripes continue to exist deep in
the superconducting phase, but only as fluctuating correlations. Such
slowly fluctuating {\it bond-aligned} incommensurate correlations have
been observed in inelastic neutron-scattering measurements of \LSCO with
$0.12\leq x \leq 0.27$\cite{wakimoto04,yamadastripesLSCO}. Their detection
along the diagonal directions would tie \LSCO with another member
of the cuprate family, \YBCO, in which an anisotropic incommensurate ring
was recently found by inelastic neutron-scattering experiments on underdoped
and nearly optimally doped samples\cite{ringYBCO}.

\acknowledgements
This research was supported by the Israel Science Foundation
(grant No. 193/02-1).

\appendix
\section{The RG Procedure for the ladder models}

\subsection{Cooper- and forward-scattering}

As mentioned previously, spin and momentum conservation generically allows
only for Cooper- and forward-scattering\cite{bfnll}. These processes are
conveniently described using the charge and spin currents
\begin{eqnarray}
J^{R}_{ij} &=& \sum_{\alpha} :\psi^{R\dag}_{i\alpha}(x)
\psi^{R}_{j\alpha}(x): \; , \nonumber \\
\bbox{J}^{R}_{ij} &=& \frac12 \sum_{\alpha,\beta} :\psi^{R\dag}_{i\alpha}(x)
\bbox{\sigma}_{\alpha \beta} \psi^{R}_{j\beta}(x): \; ,
\label{current_defs}
\end{eqnarray}
where $\bbox{\sigma}$ denotes the Pauli matrices. The same
definitions apply for the left components. The allowed terms are
\begin{eqnarray}
  H_{int} &=& \int \! dx \sum_{i,j=1}^N  \pi(v_i+v_j)
  \bigg\{c^{\rho}_{ij} J^{R}_{ij}
  J^{L}_{ij} - c^{\sigma}_{ij} \bbox{J}^{R}_{ij} \cdot
  \bbox{J}^{L}_{ij} \nonumber\\
  &+&f^{\rho}_{ij} J^{R}_{ii} J^{L}_{jj} -
  f^{\sigma}_{ij} \bbox{J}^{R}_{ii} \cdot
  \bbox{J}^{L}_{jj}\bigg\} \; ,
  \label{H_int}
\end{eqnarray}
where $v_i$ is the Fermi velocity at $k_{F_i}$.
$f_{ij}$ and $c_{ij}$ are the forward- and
Cooper-scattering amplitudes respectively. As $f_{ii}$ and
$c_{ii}$ describe the same process we take $f_{ii}=0$.

The RG equations for these couplings and their bare values in terms of $U$
and the transformation matrix $S$ introduced in Eq. (\ref{transformation})
have been derived by Lin, Balents and Fisher in Ref. \ref{bfnllref}.
The most diverging couplings, as found by numerical integration of the
RG equations, determine the nature of the phase on the scale at which the
divergence occur. Subsequent scaling-dimension analysis determines the
ultimate character of the system on even longer scales.
The precise point at which a particular operator is declared most
relevant is somewhat arbitrary. This uncertainty might affect the extent
of the phases identified. Notwithstanding, we have verified the robustness
of our results with respect to small variations in this criterion.
Note that we chose not to display in the phase diagrams, Figs.
\ref{bond2LLFig}-\ref{diag3LLFig}, some of the phases whose ranges
were found to be very small.

When none of the couplings diverge the result is a gapless
phase. The scaling-dimension analysis indicates that all the
couplings are marginal, thus ensuring the stability of the
resulting C$n$S$n$ phase (where $n$ denotes the number of pairs of
Fermi points).

A C$n$S$n${\scriptsize -1} phase is obtained at the end of the
first renormalization stage when a single $c^{\sigma}_{aa}$
coupling diverges. The bosonized relevant interaction is
proportional to
$c^{\sigma}_{aa}\cos(\sqrt{8\pi}\phi_{a}^{\sigma})$ and results in
a single gapped spin mode. A scaling-dimension analysis, after
freezing $\phi_{a}^{\sigma}$, indicates that the interactions
$c^{\sigma}_{aj}\cos[\sqrt{2\pi}(\theta_{a}^{\rho}-\theta_{j}^{\rho})]
\cos(\sqrt{2\pi}\phi_{j}^{\sigma})$ become relevant. They lead to
a new C1S0 fixed point.

The maximally allowed gapped phase, C1S0, is reached at the end of
the first RG stage when all couplings except $f^{\sigma}_{ij}$
diverge. When this happens $c_{ij}^\sigma(l^*)\approx4 c_{ij}^\rho(l^*)$ and
the relevant interactions take the form
\begin{eqnarray}
\propto \sum_{i,j}&&(v_i+v_j)
\nonumber\\
&&\times c^{\sigma}_{ij}\cos[\sqrt{2\pi}(\theta_{i}^{\rho}-\theta_{j}^{\rho})]
\cos(\sqrt{2\pi}\phi_{i}^{\sigma}) \cos(\sqrt{2\pi}\phi_{j}^{\sigma})
\; .
\nonumber \\
\label{relevant}
\end{eqnarray}
The terms with $i=j$ pin all the spin modes. The terms with $i\neq j$
imply $n-1$ gapped charge modes.
The outcome is correspondingly a C1S0 phase.

In the region where three pairs of Fermi points exist we encounter
a situation in which all the couplings associated with two Fermi point
pairs ($a,b$) diverge at the end of the first RG stage, except for
$f^{\sigma}_{ab}$. The relevant interactions are then similar to
those of Eq. (\ref{relevant}), where $i$ and $j$ take the values
$a,b$.
The $i=j$ terms lead to two gapped spin modes. The $i\neq j$ terms
result in one gapped charge mode which is associated with
$\theta_{ab}^{\rho-}=(\theta_a^{\rho}-\theta_b^{\rho})/\sqrt{2}$.

\begin{figure}[h]
\includegraphics[angle=0,width=3.2in]{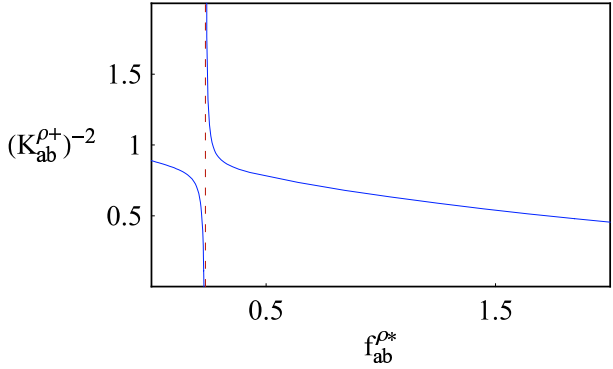}
\caption{Typical behavior of the Luttinger exponent of
$\theta_{ab}^{\rho +}$ when $\theta_{ab}^{\rho -}$ is frozen.}
\label{Kbehavior}
\end{figure}

The stability of the resulting C2S1 phase is determined by the value of
$K_{ab}^{\rho +}$, the Luttinger exponent
of $\theta_{ab}^{\rho +}=(\theta_a^{\rho}+\theta_b^{\rho})/\sqrt{2}$.
When $1/K_{ab}^{\rho +}<4$ the interactions associated with
$c_{ac}^{\sigma}$ and $c_{bc}^{\sigma}$ become
relevant and the system ends up once again in the C1S0 phase.
The value of this Luttinger exponent\cite{bf2ll}
\begin{eqnarray}
&&(K_{ab}^{\rho +})^{-2} =\\
&&\frac{4\nu[4\nu^2-4\nu(1+\nu)^2 f_{ab}^{\rho *}+(\nu^2-1)^2
(f_{ab}^{\rho *})^2]}{(1+\nu)^2[2\nu+(\nu-1)^2 f_{ab}^{\rho *}]
[2\nu-(1+6\nu+\nu^2)f_{ab}^{\rho *}]} \; , \nonumber
\end{eqnarray}
where $\nu=v_a/v_b$, depends on $f_{ab}^{\rho *}$, the forward-scattering
amplitude between the spin-gapped bands at the end of the first RG stage,
as shown in Fig. \ref{Kbehavior}. One can see that $1/K_{ab}^{\rho +}<1$,
except for a narrow region whose location varies from 1/4 to 0 as $\nu$
deviates from 1. Therefore, although the exact value of $f_{ab}^{\rho *}$
is not known precisely, we may conclude that the system eventually flows
to a C1S0 phase.

\subsection{Umklapp and ``special'' processes}

Umklapp scattering and ``special'' processes which preserve momentum
but exist only for specific values of the Fermi wave-vectors\cite{noteint} 
are allowed on special lines in parameter space, as shown, for example, 
in the phase diagrams of the 2-leg ladders. In the following we describe 
all the Umklapp and ``special'' terms in the presence of two pairs of 
Fermi points. For this purpose we define
\begin{equation}
K_{i\alpha j\beta}^{R} =
:\psi^{R}_{i\alpha}(x)\psi^{R}_{j\beta}(x): \; .
\end{equation}

An intra-band Umklapp term exists in the case where one of the Fermi points
pairs resides at $k_{F_i}=\pi/2$,
\begin{eqnarray}
 H_{Umklapp}^{(1)} = \int \! dx \sum_{\alpha,\beta} 2\pi v_i u_{ii}
\left[K_{i\alpha
 i\beta}^{R\dagger} K_{i\alpha i\beta}^{L}+h.c.\right] \, .
  \label{H_umklapp1}
\end{eqnarray}
The RG equation for the single-band Cooper-scattering
$c_{ii}^{\rho}$ is altered by this Umklapp process. Denoting the
addition to the RG equation as $\delta c_{ii}$ we find
\begin{equation}
\delta\dot{c}^{\rho}_{ii}=4(u_{ii})^2 \; ,
\end{equation}
where the dot indicates a derivative with respect to the length-scale $l$.
An additional RG equation for the Umklapp coupling itself is needed,
\begin{equation}
\dot{u}_{ii}=4c^{\rho}_{ii}u_{ii} \; ,
\end{equation}
with the initial value $u_{ii}=U/8\pi v_i$ for the systems that we
have considered.

When the sum of $k_{F_i}$ and $k_{F_j}$ at the right moving Fermi points 
equals $\pi$, three inter-pair Umklapp interactions emerge. 
\begin{eqnarray}
H_{Umklapp}^{(2)} &=&\int \! dx \sum_{\alpha,\beta}\pi(v_i+v_j) \nonumber \\
  &\times& \bigg\{u_{ij}^{a}\left[K_{i\alpha i\beta}^{R\dagger}
   K_{j\alpha j\beta}^{L}+K_{j\alpha j\beta}^{R\dagger}
   K_{i\alpha i\beta}^{L}+h.c\right] \nonumber\\
  &+&u_{ij}^{b}\left[K_{i\alpha j\beta}^{R\dagger}
   K_{i\alpha j\beta}^{L}+h.c.\right]\nonumber\\
 &-& u_{ij}^{c}\left[K_{i\alpha j\beta}^{R\dagger}
   K_{j\alpha i\beta}^{L}+h.c.\right]\:+\:i\leftrightarrow j\bigg\} \; .
  \label{H_umklapp2}
\end{eqnarray}
The same scattering processes also exist, as momentum-conserving 
``special terms'', when the two wave-vectors equal each other.  
These interactions induce additional terms in the Cooper- and
forward-scattering RG equations
\begin{eqnarray}
\delta\dot{c}^{\rho}_{ii}&=&4 \alpha_{ij}(u_{ij}^{b})^2 + 4
\alpha_{ij}(u_{ij}^{c})^2 + 4 \alpha_{ij}u_{ij}^{b}u_{ij}^{c} \; , \nonumber\\
\delta\dot{c}^{\sigma}_{ii}&=&-16 \alpha_{ij}(u_{ij}^{b})^2 - 16
 \alpha_{ij}u_{ij}^{b}u_{ij}^{c} \; , \nonumber\\
\delta\dot{f}^{\rho}_{ij}&=&16 (u_{ij}^{a})^2 + 4
(u_{ij}^{b})^2 +4 (u_{ij}^{c})^2 + 4 u_{ij}^{b}u_{ij}^{c} \; , \nonumber\\
\delta\dot{f}^{\sigma}_{ij}&=&-16 (u_{ij}^{c})^2 - 16
 u_{ij}^{b}u_{ij}^{c} \; , \nonumber\\
\delta\dot{c}^{\rho}_{ij}&=&8 u_{ij}^{a} (u_{ij}^{b} -
 u_{ij}^{c}) \; , \nonumber\\
\delta\dot{c}^{\sigma}_{ij}&=&32 u_{ij}^{a} (u_{ij}^{b} +
u_{ij}^{c}) \; ,
\end{eqnarray}
where $\alpha_{ij}=(v_i+v_j)^2/4v_iv_j$. The RG equations for the
new couplings are
\begin{eqnarray}
\dot{u}_{ij}^{a}&=&4f_{ij}^{\rho}u_{ij}^{a} +
c_{ij}^{\rho}(u_{ij}^{b}-u_{ij}^{c}) +
\frac{3}{4}c_{ij}^{\sigma}(u_{ij}^{b}+u_{ij}^{c}) \; ,\nonumber\\
\dot{u}_{ij}^{b} &=&
 u_{ij}^{a}(4c_{ij}^{\rho}+c_{ij}^{\sigma}) +
 u_{ij}^{b}(c_{ii}^{\rho}+c_{jj}^{\rho}) -
 \frac{3}{4}u_{ij}^{b}(c_{ii}^{\sigma}+c_{jj}^{\sigma}) \nonumber\\
&+& u_{ij}^{b}(2f_{ij}^{\rho}+\frac{1}{2}f_{ij}^{\sigma}) -
 \frac{1}{2}u_{ij}^{c}(c_{ii}^{\sigma}+c_{jj}^{\sigma}) \; ,
 \nonumber\\
\dot{u}_{ij}^{c} &=&
 u_{ij}^{a}(-4c_{ij}^{\rho}+c_{ij}^{\sigma}) -
 u_{ij}^{b}f_{ij}^{\sigma} +
 u_{ij}^{c}(c_{ii}^{\rho}+c_{jj}^{\rho}) \nonumber\\
&+& \frac{1}{4}u_{ij}^{c}(c_{ii}^{\sigma}+c_{jj}^{\sigma}) +
 2u_{ij}^{c}f_{ij}^{\rho} -
 \frac{3}{2}u_{ij}^{c}f_{ij}^{\sigma} \; ,
\end{eqnarray}
with the initial values $u_{ij}^{a}=U/8\pi (v_i+v_j)$,
$u_{ij}^{b}=U/4\pi (v_i+v_j)$, $u_{ij}^{c}=-U/4\pi (v_i+v_j)$.

More inter-pair Umklapp processes exist when the difference between $k_{F_i}$
and $k_{F_j}$ at the right moving Fermi points equals $\pi$,
\begin{equation}
  H_{Umklapp}^{(3)} =  \int \! dx \pi(v_i+v_j)\Big[u^{\rho}_{ij}
  J^{R}_{ij} J^{L}_{ji} + u^{\sigma}_{ij}
  \bbox{J}^{R}_{ij} \cdot \bbox{J}^{L}_{ji}\:+\:i\! \leftrightarrow
   \! j\Big] .
  \label{H_umklapp3}
\end{equation}
The additions to the Cooper- and forward-scattering RG equations are in this
case
\begin{eqnarray}
\delta\dot{c}^{\rho}_{ii}&=&\alpha_{ij}(u_{ij}^{\rho})^2 +
\frac{3}{16}\alpha_{ij}(u_{ij}^{\sigma})^2 \; , \nonumber\\
\delta\dot{c}^{\sigma}_{ii}&=&-2\alpha_{ij}u_{ij}^{\rho}u_{ij}^{\sigma}
 - \frac{1}{2} \alpha_{ij}(u_{ij}^{\sigma})^2 \; , \nonumber\\
\delta\dot{f}^{\rho}_{ij}&=&-(u_{ij}^{\rho})^2 -\frac{3}{16}
(u_{ij}^{\sigma})^2 \; , \nonumber\\
\delta\dot{f}^{\sigma}_{ij}&=&2u_{ij}^{\rho}u_{ij}^{\sigma} -
\frac{1}{2}(u_{ij}^{\sigma})^2 \; ,
\end{eqnarray}
and the RG equations for these Umklapp couplings are:
\begin{eqnarray}
\dot{u}_{ij}^{\rho}&=&u_{ij}^{\rho}(c_{ii}^{\rho}
+c_{jj}^{\rho}-2f_{ij}^{\rho})-
\frac{3}{16}u_{ij}^{\sigma}(c_{ii}^{\sigma}+c_{jj}^{\sigma}-2f_{ij}^{\sigma})
\; , \nonumber\\
\dot{u}_{ij}^{\sigma}&=&
-u_{ij}^{\rho}(c_{ii}^{\sigma}+c_{jj}^{\sigma}-2f_{ij}^{\sigma})+
u_{ij}^{\sigma}(c_{ii}^{\rho}+c_{jj}^{\rho}-2f_{ij}^{\rho})\nonumber\\
&-& \frac{1}{2}u_{ij}^{\sigma}(c_{ii}^{\sigma}
+c_{jj}^{\sigma}+2f_{ij}^{\sigma}) \; ,
\end{eqnarray}
with the initial values $u_{ij}^{\rho}=U/4\pi (v_i+v_j)$,
$u_{ij}^{\sigma}=-U/\pi (v_i+v_j)$.

In the diagonal 2-leg ladder, interactions that involve an odd
number of operators per band are allowed on specific lines. For example, when
the sum of $k_{F_i}$ and $3k_{F_j}$ at the right moving Fermi
points equals zero or equals $2\pi$, we find:
\begin{eqnarray}
  H_{Umklapp}^{(4)} &=&  \int \! dx \sum_{\alpha,\beta}\pi(v_i+v_j)
  u^{d}_{ij}
  \Big[ K_{i\alpha j\beta}^{R\dagger} K_{j\alpha j\beta}^{L} \nonumber \\
  &+& K_{i\alpha j\beta}^{L\dagger} K_{j\alpha j\beta}^{R}+h.c \Big] \; .
  \label{H_umklapp4}
\end{eqnarray}
This coupling induces the following modifications to the RG flow equations 
\begin{eqnarray}
\delta\dot{c}^{\rho}_{jj}&=&2 \beta_{ij}(u_{ij}^{d})^2 \; , \nonumber\\
\delta\dot{c}^{\rho}_{ij}&=& \beta_{ij}(u_{ij}^{d})^2  \; , \nonumber\\
\delta\dot{f}^{\rho}_{ij}&=& \beta_{ij}(u_{ij}^{d})^2 \; ,
\end{eqnarray}
where $\beta_{ij}=(v_i+v_j)/2v_j$. The additional RG equation is
\begin{equation}
\dot{u}_{ij}^{d}=2u_{ij}^{d}\left(c^{\rho}_{jj} + c^{\rho}_{ij} +
f^{\rho}_{ij}\right) \; ,
\end{equation}
with the initial value $u_{ij}^{d}=U/2\pi (v_i+v_j)$.

In the case that the difference between $3k_{F_i}$ and $k_{F_j}$
at the right moving Fermi points equals $2\pi$, the Hamiltonian
is:
\begin{eqnarray}
  H_{Umklapp}^{(5)}\! &=&\!  \int \! dx \! \sum_{\alpha,\beta}
  \pi(v_i+v_j)u^{e}_{ij}
  \Big[ K_{i\alpha i\beta}^{R\dagger} 
  :\!\psi^{L}_{i\alpha}(x)\psi^{R}_{j\beta}(x)\!:
  \nonumber \\&+&
 K_{i\alpha i\beta}^{L\dagger} :\!\psi^{R}_{i\alpha}(x)\psi^{L}_{j\beta}(x)\! :
 +h.c \Big] \; .
  \label{H_umklapp5}
\end{eqnarray}
The additions to the Cooper- and forward-scattering RG
equations in this case are
\begin{eqnarray}
\delta\dot{c}^{\rho}_{ij}&=& -\frac{1}{2} \beta_{ji}(u_{ij}^{e})^2  \; , 
\nonumber\\
\delta\dot{c}^{\sigma}_{ij}&=& 2 \beta_{ji}(u_{ij}^{e})^2 \; .
\end{eqnarray}
The RG equation for the new coupling is
\begin{equation}
\dot{u}_{ij}^{e}= u_{ij}^{e}\left(2 c^{\rho}_{ii} - f^{\rho}_{ij}
- f^{\sigma}_{ij}\right) \; ,
\end{equation}
with the initial value $u_{ij}^{e}=U/2\pi (v_i+v_j)$.

\end{multicols}


\begin{references}
\bibitem{comment} In some of these studies the second RG stage was not
carried out. This omission sometimes led to the conclusion that the system
remains gapless, where in fact it ultimately reaches the C1S0 phase.
\bibitem{noteint} Note that the distinction between Umklapp and ``special'' 
processes depends on the choice of unit cell and is therefore somewhat 
ambiguous.
\end{references}
\end{document}